\documentclass[10pt,notitlepage,final]{iopart}

\usepackage{iopams}  
\usepackage{hyperref}
\usepackage{graphicx}
\usepackage{multirow}

\newcommand{\be}{\begin{equation}}
\newcommand{\ee}{  \end{equation}}
\newcommand{\ba}{\begin{eqnarray}}
\newcommand{\ea}{  \end{eqnarray}}
\newcommand{\ve}{\varepsilon}

\begin{document}

\title{Effective field theory for deformed atomic nuclei}

\author{T.~Papenbrock$^{1,2}$ and H.~A. Weidenm{\"u}ller$^{3}$}

\address{$^1$Department of Physics and Astronomy, University of
  Tennessee, Knoxville, TN 37996, USA}

\address{$^2$Physics Division, Oak Ridge National Laboratory, Oak
  Ridge, TN 37831 USA}

\address{$^3$Max-Planck Institut f\"ur Kernphysik, D-69029
  Heidelberg, Germany}

\eads{\mailto{tpapenbr@utk.edu}, \mailto{haw@mpi-hd.mpg.de}}

\begin{abstract}
  We present an effective field theory (EFT) for a model-independent
  description of deformed atomic nuclei. In leading order this
  approach recovers the well-known results from the collective model
  by Bohr and Mottelson. When higher-order corrections are computed,
  the EFT accounts for finer details such as the variation of the
  moment of inertia with the band head and the small magnitudes of
  interband $E2$ transitions. For rotational bands with a finite spin
  of the band head, the EFT is equivalent to the theory of a charged
  particle on the sphere subject to a magnetic monopole field.
\end{abstract}

\submitted{\PS}

\maketitle


\section{Introduction}
\label{int}

The ground-breaking papers by Bohr~\cite{bohr_1952} and by Bohr and
Mottelson~\cite{bohrmottelson_1953} have laid the foundation for our
understanding of collective motion in heavy nuclei~\cite{bmbook}. In
that approach, collective excitations are quantized surface
oscillations of a liquid drop, with quadrupole modes dominating at low
energies. The five degrees of freedom associated with quadrupole
deformations of a spherical surface are chosen as the three Euler
angles (that describe the orientation of the nonspherical liquid drop)
and two shape parameters $\beta$ and $\gamma$ (that describe the
amplitudes of the deformation in the body-fixed, i.e. the co-rotating,
coordinate system of the liquid drop). The parameter $\beta$ describes
the deformation of an axially symmetric drop, while $\gamma$ describes
deformations that break the axial symmetry.

The Bohr Hamiltonian governs the dynamics of the collective degrees of
freedom. The kinetic term is that of a five-dimensional harmonic
quadrupole oscillator. Because of rotational invariance, the potential
term is a function of the two shape parameters $\beta$ and $\gamma$
alone. The recent compendium~\cite{fortunato2005} lists and describes
analytical solutions known for special cases. For general potentials,
the Bohr Hamiltonian must be solved numerically. Powerful numerical
methods have been developed for this purpose only
recently~\cite{rowe2004}.

Low-lying spectra of deformed nuclei are characterized by three energy
scales. The lowest excitations are quantized rotations. In deformed
rare-earth nuclei (in actinides) the associated energy scale $\xi$ is
$\xi \approx 80$~keV ($\xi\approx 40$~keV, respectively). The spectra
consist of rotational bands built on top of vibrational band heads;
the latter are excitations of the $\beta$ and $\gamma$ degrees of
freedom. The energy scale $\Omega$ associated with vibrational
excitations is $\Omega \approx 800$~keV ($\Omega \approx 500$~keV,
respectively). The low-energy description of even-even deformed nuclei
in terms of bosonic collective modes breaks down at an energy scale
$\Lambda$ where single-particle effects that reveal the fermionic
nature of the nucleonic degrees of freedom become important. In
even-even heavy nuclei, pair breaking occurs at $\Lambda \approx
2$--3~MeV.

The scales $\xi, \Omega, \Lambda$ are separated. The inequality $\xi
\ll \Omega$ holds very well, while $\Omega \ll \Lambda$ holds
marginally. These facts suggest that a model-independent description
of deformed nuclei in terms of an effective field theory (EFT) should
be useful. An EFT has several advantages over phenomenological
models. First, the separation of scales allows for the introduction of
the small expansion parameters $\xi / \Lambda$ and $\Omega /
\Lambda$. Power counting can be used to order contributions to the
Hamiltonian. Keeping terms up to a given order yields a Hamiltonian
with quantifiable theoretical uncertainties. The Hamiltonian can be
improved and the errors reduced by including terms of the next order.
Second, the Hamiltonian and the interaction currents are treated on an
equal footing. Transition operators and observables other than the
energy can be treated consistently up to the same order as the
Hamiltonian. Third, an EFT makes it possible to establish relationships
between observables (instead of relationships between model
parameters).

We stress that an EFT approach to nuclear deformation is not of purely
theoretical interest only. The Bohr Hamiltonian and other, more
general collective models describe the gross properties of even-even
deformed nuclei very well. Examples are the rotational spectra on top
of vibrations and the strong intra-band $E2$ transitions. Finer
details are not reproduced correctly. For instance, in the collective
models the moments of inertia typically tend to decrease with
increasing vibrational excitation energies, in contrast to the
data. The magnitudes of the (weak) inter-band $E2$ transitions are
overpredicted by factors 2--10~\cite{rowe2010}.

In the present article we review recent uses of EFT for the
description of deformed nuclei. Some of
these~\cite{papenbrock2011,zhang2013,coello2015} have employed
``effective theory'' and were aimed at a direct comparison with the
data. The shortcomings of the models mentioned in the previous
paragraph have been addressed
successfully~\cite{zhang2013,coello2015}.  Other
papers~\cite{papenbrock2014,papenbrock2015} developed an EFT for
``emergent symmetry breaking'' (a general phenomenon in finite
systems), with special application to deformed nuclei. Here we aim at
a unified presentation of these approaches. We will not repeat the
detailed derivations given in the above-mentioned articles. We aim at
presenting the main physical ideas while keeping the formalism at a
minimum.

The use of EFT in nuclear physics is of course not restricted to
deformed rotating nuclei. On the contrary, in the last two decades
EFTs have been widely used. Chiral EFT exploits the separation of
scale between the long-ranged pion exchange (resulting from the
spontaneous breaking of chiral symmetry in low-energy quantum
chromodynamics) and short-ranged interactions for the systematic
construction of the inter-nucleon
potentials~\cite{vankolck1994,epelbaum2009,machleidt2011}. Interactions
and currents from chiral EFT serve as input for nuclear-structure
calculations for nuclei with up to medium
mass~\cite{navratil2009,barrett2013,hagen2014} and neutron-rich
matter~\cite{hebeler2015}. Pion-less EFT has been used to describe
few-body nuclear
systems~\cite{kaplan1998,bedaque2002,griesshammer2012,hammer2013}. Here,
the unknown short-ranged interaction is systematically parameterized
by contact terms and derivatives. Halo
EFT~\cite{bertulani2002,hammer2011,ryberg2014} exploits the separation
of scale between weakly-bound halo states and higher-lying excitations
and describes halo nuclei in terms of valence nucleons that are weakly
bound to inert core states.

The article is organized as follows. We introduce emergent symmetry
breaking in Sect.~\ref{eme}.  In Sect.~\ref{construction} we review
the construction of the EFT for deformed nuclei. Section~\ref{results}
presents results of this approach, compares the EFT to data and to
effective theories, and mentions some interesting problems that have
been addressed this way. We give a summary and brief outlook in
Sect.~\ref{summary}. Some technical details are presented in \ref{app}.

\section{Spontaneous versus emergent symmetry breaking}
\label{eme}

We speak of spontaneous symmetry breaking (both in relativistic and in
infinitely extended nonrelativistic systems) when the ground state of
a quantum system does not possess the full symmetry of the
Hamiltonian. A case in point is the infinite ferromagnet. In the
ground state, all spins point in the same direction, violating the
rotational invariance of the Hamiltonian. The ground states of two
infinite ferromagnets that differ in the orientations of their spins
have zero overlap. Indeed, that overlap is the product of the overlaps
of infinitely many individual spin states all of which are less than
unity. Therefore, a unitary transformation that would link the two
gound states, does not exist: The two Hilbert spaces built upon the
two ground states are unitarily inequivalent. A rotation is
quantum-mechanically described by a unitary transformation in Hilbert
space. Therefore, the infinite ferromagnet cannot rotate. EFT offers a
way to determine the low-lying excitations of such systems using
symmetry arguments only. In the ferromagnet, these are spin waves with
arbitrarily long wave lengths. The EFT construction uses the Goldstone
theorem.  Low-lying excitations are given in terms of the
Nambu-Goldstone modes.

Spontaneous symmetry breaking does not occur in finite systems. Ground
states of finite ferromagnets, for instance, that differ in spin
orientation have nonzero overlap and are linked by unitary
transformations describing rotations. Therefore, a finite ferromagnet
can rotate about any axis perpendicular to the symmetry axis of the
ground state, and rotational invariance is restored. However, there is
a smooth transition to the case of spontaneously broken symmetry: As
the size of the ferromagnet increases, the overlap of ground states
oriented in different directions decreases exponentially and becomes
zero as the size tends to infinity. Then rotational motion cannot
occur. We speak of emergent symmetry breaking in a finite system when
spontaneous symmetry breaking occurs in the limit of infinite system
size. Thus, emergent symmetry breaking is the precursor of spontaneous
symmetry breaking. The ground state of the finite system possesses the
full symmetry of the Hamiltonian. Nevertheless, some of the low-lying
excitations are akin to those of the infinite system. The EFT
construction is, thus, expected to apply also for emergent symmetry
breaking, augmented, of course, by degrees of freedom that account for
rotational motion.

That is the idea we pursue in the present paper. Nuclei are invariant
under rotations, and angular momentum is a good quantum number. But
heavy deformed nuclei possess an intrinsic ground state that is
deformed and, thus, breaks the rotational invariance of the
Hamiltonian in the limit of infinite system size. We consider that a
case of emergent symmetry breaking~\cite{yannouleas2007}. We modify
the EFT approach to infinite systems so as to take account of
rotational motion. The approach is expected to work best for large
systems (with large moments of inertia) as we are then close to the
infinite case. The energy scales discussed in the introduction show
that that condition is met very well in deformed nuclei. The broken
symmetry is rotational $SO(3)$, the deformed ground state is invariant
under $SO(2)$ rotations about the body-fixed symmetry axis.  Hence,
our generalized EFT is based on Nambu-Goldstone modes constructed in
the $SO(3) / SO(2)$ coset space, a nonlinear realization of the
underlying symmetry.

We compare that approach to the standard description of nuclei.
Practically all theoretical approaches to nuclei employ spherical
tensors as fundamental tools, i.e., use the Wigner-Weyl (linear)
realization of $SO(3)$ symmetry. The Bohr Hamiltonian and the general
collective models, for instance, employ quadrupole degrees of freedom,
i.e., spherical tensors of rank two, while the interacting boson model
(IBM)~\cite{iachello1987} employs in addition to the $d$ bosons also
an $s$ boson. In Bohr's original approach~\cite{bohr_1952}, the
transformation from quadrupole degrees of freedom to Euler angles and
shape parameters is from the laboratory to the body-fixed system. That
non-linear transformation is complicated, and it is not obvious how to
construct, for instance, higher-order kinetic terms directly in the
body-fixed coordinates. Moreover, in the geometric and algebraic
collective models, the description of deformed nuclei requires large
basis sets when the basis of the quadrupole oscillator is
used. Efficient basis construction schemes and formulas for many
relevant matrix elements have only been given
recently~\cite{rowe2004,rowe2009}.

In contrast to these bosonic descriptions, more microscopic approaches
such as the nuclear shell model are based on fermionic degrees of
freedom, i.e., on spherical tensors of half-integer rank. For deformed
nuclei, such microscopic approaches are complicated and/or numerically
expensive. Mean-field calculations, for instance, yield a product
state that breaks rotational invariance, and projection methods need
be employed for the restoration of rotational
symmetry~\cite{nazarewicz1994,bender2003}. Likewise, the Nilsson
model~\cite{nilsson} constructs deformed states in an intrinsic
(co-rotating) frame, and one is left with the problem to couple the
intrinsic states to rotations. {\it Ab initio} methods have recently
succeeded in describing emergent phenomena such as rotational bands in
light nuclei~\cite{caprio2013,dytrych2013}. These tour-de-force
computations employ large model spaces~\cite{barrett2013} and/or
complicated basis states~\cite{dytrych2008} for the description of
phenomena linked to rotational motion.

\section{EFT for deformed nuclei}
\label{construction}

Effective field theories are based on the symmetries and the pattern
of symmetry breaking relevant for the physical system under
consideration. For systems with spontaneous symmetry breaking,
Weinberg~\cite{weinberg1968}, Coleman {\it et al.}~\cite{coleman1969},
and Callan {\it et al.}~\cite{callan1969} devised methods for the
construction of effective Lagrangeans. These approaches are reviewed
in Refs.~\cite{leutwyler1994,weinberg_v2_1996,brauner2010}. For
deformed nuclei, we deal with emergent symmetry breaking from
rotational $SO(3)$ to axial $SO(2)$ symmetry. The EFT approach to
these finite systems is to a large extent similar to systems such as
(anti-)ferromagnets~\cite{leutwyler1994,roman1999,hofmann1999,baer2004,kampfer2005}
that also exhibit a spontaneous breaking of $SO(3)$ to $SO(2)$
symmetry, but modifications arise due to the emergent character of the
symmetry breaking.

In systems with a spontaneous breaking of the symmetry group ${\cal G}$
to a subgroup ${\cal H}$, the low-energy degrees of freedom are
Nambu-Goldstone (NG) bosons which parameterize the coset space ${\cal
  G}/{\cal H}$~\cite{coleman1969,callan1969}. In our case, ${\cal G} =
SO(3)$ and ${\cal H} = SO(2)$. The NG bosons are fields that depend on
spatial and temporal coordinates. In the infinite system, a purely
time-dependent mode (that is constant in space) is excluded because
such a mode connects unitarily inequivalent Hilbert spaces (in the
ferromagnet: symmetry axes that point in macroscopically different
directions). In finite systems with emergent symmetry breaking, that
zero mode cannot be neglected. In our case, the mode corresponds to
overall rotations of the system and connects Hilbert spaces that
become inequivalent as the system size approaches infinity. It is that
mode that restores the rotational symmetry and allows us to label
states by total angular momentum. Such zero modes are also important
for finite-size effects in numerical simulations of field theories in
a finite volume~\cite{leutwyler1987,gasser1988,hasenfratz1993}. We
summarize the construction of an EFT based on emergent symmetry
breaking for the case of deformed nuclei developed in
Refs.~\cite{papenbrock2014,papenbrock2015}.

For deformed nuclei, the relevant low-energy degrees of freedom are NG
fields $\psi_x(\mathbf{x},t)$ and $\psi_y(\mathbf{x},t)$, and
rotations are described by angles $\phi(t),\theta(t)$. Here,
$\mathbf{x}=(x_1,x_2,x_3)$ denotes the three spatial coordinates of
the fields. The NG fields and the rotation angles parameterize the
coset space $SO(3)/SO(2)$ which is isomorphic to the two-sphere
$S^2$. We consider
\ba 
\label{U}
U &\equiv & g(\phi,\theta) u(\psi_x,\psi_y)\ , \\ 
\label{g}
g(\phi,\theta) &\equiv & e^{-i\phi \hat{J}_z} e^{-i\theta \hat{J}_y} \ , \\ 
u(\psi_x,\psi_y) &=& e^{-i\psi_x\hat{J}_x-i\psi_y\hat{J}_y} \ .  
\ea 
The three angular-momentum operators $\hat{J}_k$, $k = x, y, z$
generate rotations around the $k$-axes in the space-fixed coordinate
system. They fulfill the usual commutation relations $\left[
  \hat{J}_x, \hat{J}_y\right] = i \hat{J}_z$ (and cyclic
permutations)~\footnote{We could also use operators that generate
  rotations around the body-fixed $k$-axes. Then the commutation
  relations carry different signs. That would not change the physical
  picture.}. The physical picture is this. We choose the $z$-axis of
the laboratory system such that it initially coincides with the
symmetry axis of the deformed nucleus, and we apply $U=gu$ of
Eq.~(\ref{U}). For small $\psi_x(\mathbf{x},t)$ and
$\psi_y(\mathbf{x},t)$ the operator $u$ causes position-dependent
small-angle distortions of each volume element (each nucleon) of the
liquid drop (of the nucleus consisting of nucleons, respectively).
These distortions have to be quantized. The NG fields
$\psi_x(\mathbf{x},t)$ and $\psi_y(\mathbf{x},t)$, thus, induce
quantized vibrations at the energy scale $\Omega$. The operator $u$ is
followed by $g$ which generates an overall rotation of the vibrating
nucleus. The new symmetry axis is oriented along the vector
\ba
\label{er}
\vec{e}_r \equiv \left(\begin{array}{c}
\cos \phi \sin \theta\\
\sin \phi \sin \theta\\
\cos \theta\end{array}\right) \ .
\ea
The vector $\vec{e}_r$ is defined in actual physical space. The vector
$\mathbf{x}$ refers to the coordinates of the constituents of the
nucleus. For that reason the two vectors are denoted differently. We
are going to use a small-amplitude approximation for the NG fields
$\psi_x(\mathbf{x},t)$ and $\psi_y(\mathbf{x},t)$ whereas the rotation
described by the angles $\theta$ and $\phi$ is fully taken into
account.

\subsection{Transformation properties under rotations}

To study the transformation properties of our degrees of freedom under
rotations, we act onto $U$ defined in Eq.~(\ref{U}) with a general
rotation $f(\alpha, \beta, \gamma) = \exp{(-i\alpha \hat{J_z})}
\exp{(-i\beta \hat{J_y})} \exp{(-i\gamma \hat{J_z})}$ with Euler
angles $(\alpha,\beta,\gamma)$. From Eq.~(\ref{g}) we have
\ba
f(\alpha,\beta,\gamma)&=&g(\alpha,\beta)h(\gamma) \ .
\ea
Here
\ba
h(\gamma)&=&e^{-i\gamma\hat{J}_z}
\ea
is a pure rotation around the $z$-axis. Any rotation $f$ acting on a
rotation $g$ yields a new rotation $f'$. Thus,
\ba
f(\alpha, \beta, \gamma) g(\phi, \theta) &=& f(\alpha', \beta', \gamma')
\nonumber \\
\label{g'}
&=& g(\phi' ,\theta') h(\gamma') \ .
\ea
Eq.~(\ref{g'}) defines the rotated angles $(\phi', \theta', \gamma')$.
Here $\gamma'=\gamma'(\alpha, \beta, \gamma, \phi, \theta)$ is, in
general, a complicated function of its arguments. Acting with $f$ onto
$U$ of Eq.~(\ref{U}) yields
\ba
f(\alpha,\beta,\gamma) U &=& g(\phi',\theta')h(\gamma')u(\psi_x,\psi_y) \nonumber\\
&=& g(\phi',\theta')\left[h(\gamma')u(\psi_x,\psi_y)h^{-1}(\gamma')\right] h(\gamma') \nonumber\\
&=& g(\phi',\theta')u(\psi_x',\psi_y') h(\gamma') \ .
\ea
The last line defines the rotated NG fields $\psi_x'$ and $\psi_y'$
through
\be
u(\psi_x',\psi_y') \equiv h(\gamma')u(\psi_x,\psi_y)h^{-1}(\gamma') \ .
\ee
The nontrivial result of these transformation properties is that the
NG fields transform linearly (albeit with a complicated transformation
angle $\gamma'$), while the rotation angles $(\phi,\theta)$ transform
nonlinearly. Equation~(\ref{er}) shows that the rotation angles
$(\phi,\theta)$ can be viewed as azimuthal and polar angles that
parameterize the two-sphere; under rotations, they transform
accordingly. The NG fields can be viewed as parameterizing the tangent
plane of the two-sphere at $(\phi,\theta)$; under rotations they
transform as the components of a vector in the tangent plane, i.e.,
they undergo an $SO(2)$ rotation about the angle $\gamma'$ around an
axis perpendicular to the tangent plane.

In our parameterization the limit to the infinite system is not
smooth. In the infinite system $(\phi,\theta)$ assume constant (i.e.,
time-independent) values, and we can for simplicity set $g = 1$. Then
under rotations $fU = fgu = fu$, and the NG fields $(\psi_x, \psi_y)$
transform non-linearly. Details about the transformation properties of
the EFT are presented in the~\ref{app}.

\subsection{Effective Lagrangean}

The Lagrangean must be invariant under rotations. It is obtained as a
linear combination of invariants that are formed from the NG fields
and their derivatives with respect to time and space. The coefficients
of the linear combination are the parameters of the theory. These must
be adjusted to the data. The full construction, extensively discussed
in
Refs.~\cite{papenbrock2011,zhang2013,papenbrock2014,coello2015,papenbrock2015},
shows that invariants involving neither spatial nor temporal
derivatives do not occur. The invariants are ordered using the energy
scales introduced in Section~\ref{eme}. The resulting power-counting
scheme is similar to the one used in Section~\ref{pow} below for the
Hamiltonian and is not discussed here, see
Refs.~\cite{papenbrock2014,papenbrock2015}. The resulting Lagrangean
(containing, for instance, only leading-order or
leading-plus-next-to-leading-order terms) defines a classical field
theory. It has (at least) two constants of motion, the energy and the
total angular momentum of the system.

In the present Section we do not follow the procedure of
Refs.~\cite{papenbrock2014,papenbrock2015} for the construction of the
invariants but employ a shortcut suggested by the geometric picture
given above. As the coset space $SO(3)/SO(2)$ is isomorphic to the
two-sphere, the dynamics of rotations is that of a particle on that
sphere. The rotation angles $\theta$ and $\phi$ appear as arguments of
the radial vector
\be
\label{vecr}
\vec{r}=r\vec{e}_r
\ee
that parameterizes the surface of the two-sphere sphere of radius $r$.
The parameter $r$ is a low-energy constant (LEC) to be determined
later. The velocity of the particle is
\be
\label{vecv}
\vec{v} \equiv r \partial_t \vec{e}_r  = r \dot{\theta} \vec{e}_\theta +
r\dot{\phi} \sin\theta \vec{e}_\phi \ .
\ee
Here, the dot over a variable denotes its time derivative. The vector
$\vec{v}$ lies in the plane tangent to the sphere at the point
$(\phi,\theta)$. The plane is spanned by the azimuthal and polar unit
vectors
\ba
\vec{e}_\theta \equiv \left(\begin{array}{c}
\cos\phi \cos \theta\\
\sin\phi \cos \theta\\
-\sin \theta \end{array} \right) \ {\rm and} \quad 
\vec{e}_\phi \equiv \left(\begin{array}{c}
-\sin \phi\\
\cos \phi \\
0\end{array} \right) \ .
\ea 
Under rotations, the components $(v_\theta, v_\phi) = (r \dot{\theta},
r \dot{\phi} \sin \theta)$ of the velocity vector transform in the
tangential plane as do the NG fields $(\psi_x, \psi_y)$.

For the time derivatives of the NG fields we use the covariant
derivative
\be
\label{Dt}
D_t = \partial_t -i \dot{\phi} \cos \theta \hat{J}_z \ .
\ee
The operator $D_t$ projects the time derivative of a vector in the
tangential plane back onto the tangential plane. When $D_t$ acts on
the NG fields $(\psi_x, \psi_y)$, the operator $\hat{J}_z$ in
Eq.~(\ref{Dt}) acts as $- i (\psi_x \partial \psi_y - \psi_y
\partial \psi_x)$. That yields
\ba
D_t\left(\begin{array}{c}
\psi_x\\
\psi_y\end{array}\right) = \left(\begin{array}{c}
\dot{\psi}_x\\
\dot{\psi}_y\end{array}\right) +
\dot{\phi}\cos\theta\left(\begin{array}{c}
\psi_y\\
-\psi_x\end{array}\right) \ .
\ea

Spatial derivatives of the NG fields are written as $\partial_k
\psi_x(\mathbf{x}, t)$ or $\partial_k \psi_y(\mathbf{x}, t)$ with $k =
1, 2, 3$. Here $\partial_k\equiv {\partial /\partial x_k}$ denotes the
partial derivative with respect to the component $x_k$ of the
coordinate vector $\mathbf{x}=(x_1, x_2, x_3)$. As is the case for the
NG fields themselves, for each point $\mathbf{x}$ the derivative
fields lie in the plane tangent to the two-sphere at $(\phi, \theta)$
and under rotations transform accordingly. For a coordinate system
chosen such that the symmetry axis of the deformed nucleus is
initially along $\mathbf{x} = (0, 0, x_3)$, combinations of
derivatives that respect that symmetry such as $\partial_1^2 +
\partial_2^2$ or $\partial_3^2$ are permitted.

In summary, scalars constructed from the vectors $(\dot{\theta},
\dot{\phi} \sin\theta)$, $(D_t \psi_x, D_t \psi_y)$, and $\partial_k
(\psi_x, \psi_y)$ (that all lie within the tangential plane) that are
invariant under $SO(2)$ rotations are formally invariant under $SO(3)$
rotations and form the building blocks of the effective Lagrangean.
Invariants such as $(\psi_x^2 + \psi_y^2)$ that involve only the NG
vectors $(\psi_x, \psi_y)$ and not their derivatives are not allowed.
However, invariants of that type may be used as factors multiplying
any of the scalars formed from the derivatives.

As for discrete symmetries, we consider parity, ${\cal R}$ parity, and
time-reversal. The fields $\psi_x$ and $\psi_y$ have positive parity
because the corresponding generators $\hat{J}_x$ and $\hat{J}_y$ do
[see Eq.~(\ref{U})]. Thus, all states that result from the EFT
construction have positive parity. ${\cal R}$ parity is defined as the
result of a rotation by $\pi$ around any axis perpendicular to the
axis of axial symmetry. In nuclei that operation maps the ground state
onto itself, so that nuclei have have positive ${\cal R}$
parity~\cite{bmbook}. That fact distinguishes nuclei from molecules.
Time-reversal invariance restricts invariants to terms that contain an
even number of time derivatives.

The effective Lagrangean obtained that way is rotationally invariant.
That fact implies that the total angular momentum $\vec{I}$ is
conserved. It is given by (see \ref{app} for details)
\ba
\label{I}
\vec{I}&=&p_\theta\vec{e}_\phi -{p_\phi-K\cos\theta\over\sin\theta}\vec{e}_\theta +K\vec{e}_r \ .
\ea
Here $p_\phi$ and $p_\theta$ are the canonical momenta of $\phi$ and
$\theta$, respectively, defined as usual in terms of partial
derivatives of the Lagrangean. The contribution to the angular
momentum of the NG fields is
\be 
\label{K}
K \equiv \int {\rm d}^3x (\psi_x p_y - \psi_y p_x) \ , 
\ee 
where $p_x$ and $p_y$ denote the canonical momenta of the NG fields
$\psi_x$ and $\psi_y$, respectively. For every point $\mathbf{x}$ of
the fields, the term $(\psi_x p_y - \psi_y p_x)$ is a scalar under
rotations. Hence $K$ is a scalar as well. The square of the total
angular momentum is
\be
I^2 = p_\theta^2 +{1\over\sin^2\theta}\left(p_\phi^2 - 2K p_\phi \cos\theta +K^2\right) \ .
\ee
Eqs.~(\ref{I}) and (\ref{K}) follow from an application of the Noether
theorem to the effective Lagrangean, see
Refs.~\cite{papenbrock2014,papenbrock2015}.

\subsection{Effective Classical Hamiltonian}

In Refs.~\cite{papenbrock2014,papenbrock2015}, a Legendre
transformation of the effective Lagrangean leads to the effective
Hamiltonian, which in turn is quantized and employed for the
computation of spectra and transitions. That procedure is
straightforward but tedious, especially when terms beyond leading
order are considered. In the present Section we show how an effective
Hamiltonian that is invariant under rotations may be constructed
directly.

\subsubsection{Kinetic Terms.}
\label{kin}

The Legendre transformation effectively replaces time derivatives of
variables by their canonical momenta. Spatial derivatives
$\partial_k(\psi_x, \psi_y)$ are not affected by the transformation.
Therefore, we first address the kinetic terms in the effective
Hamiltonian.

The vector
\be
\label{pvec}
\vec{p} = {p_\theta \over r} \vec{e}_\theta + {p_\phi \over r \sin \theta}
\vec{e}_\phi
\ee
denotes the momentum of a particle on the two-sphere with radius $r$.
It is conjugate to the vector $\vec{r}$ in Eq.~(\ref{vecr}). The
vector $\vec{p}$ lies in the tangent plane of the two-sphere at the
point $(\theta, \phi)$. The components $p_\theta / r$ and $p_\phi / (r
\sin \theta)$ transform accordingly. The momenta of the NG fields
$(p_x, p_y)$ are likewise components of a vector in the tangent plane
and under rotations transform accordingly as well. Kinetic terms in
the Hamiltonian that are invariant under rotations are, thus, scalars
in the tangential plane.

Rotations and surface vibrations are coupled in a subtle way. That is
seen by rewriting the angular momentum in Eq.~(\ref{I}) as
\be
\label{Ialt}
\vec{I} = \vec{r} \times \left(\vec{p}-{K\cot\theta\over r} \vec{e}_\phi\right) + K\vec{e}_r \ .
\ee
The term $K\vec{e}_r$ is the intrinsic angular momentum along the
symmetry axis of the deformed nucleus. The cross product in
Eq.~(\ref{Ialt}) is the angular momentum associated with the rotor
(i.e. with a particle on the sphere of radius $r$). The vector
\be 
\label{gauge}
\vec{p} - {K\cot\theta\over r}\vec{e}_\phi \ .  
\ee
may be viewed as the ``gauged'' momentum of a particle on the sphere
subject to a vector potential
\be
\label{Amono}
\vec{A}\equiv {K\cot\theta\over r}\vec{e}_\phi
\ee
with associated magnetic field
\ba
\vec{B}&=&\vec{\nabla}\times \vec{A} \nonumber\\
&=&\left(\vec{e}_r \partial_r +\vec{e}_\theta {1\over r}\partial_\theta +\vec{e}_\phi{1\over r\sin\theta}\partial_\phi\right)\times\vec{A}\nonumber\\
&=&-{K\over r^2}\vec{e}_r \ .
\ea
The field $\vec{B}$ is normal to the surface of the sphere. It is
obviously invariant under rotations. We note that
\be
\label{moment}
{1\over 2}\left(\vec{p} - {K\cot\theta\over r}\vec{e}_\phi \right)^2
= {\vec{I}^2 - K^2 \over 2 r^2}
\ee
is the Hamiltonian of an axially symmetric rotor, with $r^2$ the
moment of inertia. We also note that rotations change the vector
potential~(\ref{Amono}) by a gradient of a scalar function. Thus,
rotations induce an unobservable gauge transformation but otherwise
leave gauged kinetic terms invariant. Details are presented
in~\ref{app}.

Vibrational and rotational degrees of freedom are, thus, coupled by
the gauge field $\vec{A}$. The vector potential~(\ref{Amono}) is that
of a magnetic monopole with charge $K$ at the center of the
sphere~\cite{wu1976,chandrasekharan2008}. (For a monopole it would be
required that the ``charge'' $K$ is integer or half integer.)
Alternatively we may say that the intrinsic angular momentum
$K\vec{e}_r$ generates a magnetic moment. The resulting magnetic field
at the point $\vec{r}$ (which points in the radial direction) couples
vibrational and rotational motion.

This insight allows one develop an EFT for rotational bands with a
finite spin $S$ of the band head, based on rotational degrees of
freedom $(\theta,\phi)$ alone (i.e. without introduction of
vibrational degrees of freedom). We note that odd-mass nuclei and
odd-odd nuclei have finite spins in their ground states.  The
Hamiltonian
\be
\label{odd}
{1\over 2}\left(\vec{p} - {S\cot\theta\over r}\vec{e}_\phi \right)^2
= {\vec{I}^2 - S^2 \over 2 r^2}
\ee
with $S$ being an integer or a half-integer constant is clearly
invariant under rotations. For $|S|>1/2$, this Hamiltonian indeed
yields the leading-order description of nuclei with a ground-state
spin $S$~\cite{bmbook}. Two comments are in order. Let us first
consider time-reversal invariance. Under time reversal $\vec{p}\to
-\vec{p}$, and $K\to -K$, but $S$ is a constant. This confirms that
the Hamiltonian~(\ref{moment}) is invariant under time reversal (it
was constructed this way) in contrast to the
Hamiltonian~(\ref{odd}). This explains why the latter could not
directly be derived within the EFT presented in this paper. Second,
Coriolis forces modify the spectrum of the Hamiltonian~(\ref{odd}) at
leading order for spins $S=1/2$.  We remind the reader that Coriolis
forces couple the angular momentum of the rotor to the nuclear spin
$S$, and they can induce spin flips. Such spin flips correspond to
$S\to -S$ and are only possible if $S$ is a dynamical degree of
freedom (and not a constant). Thus, the Hamiltonian~(\ref{odd})
describes the leading-order physics only for $|S|>1/2$, because only
then are spin-flips higher energetic excitations and beyond leading
order. For a full-fledged EFT of deformed odd-mass nuclei, one would
need to couple nucleons to the rotor.

\subsubsection{Power counting.}
\label{pow}

Power counting is based on the scales $\xi$ and $\Omega$ defined in
Section~\ref{eme} and on the relations
\ba
\label{power}
\phi, \theta &\sim& {\cal O}(1) \ , \nonumber \\
\dot{\phi}, \dot{\theta} &\sim& \xi \ , \nonumber \\
p_\phi, p_\theta &\sim& {\cal O}(1) \ , \nonumber \\
\psi_x, \psi_y &\sim& \sqrt{\xi/\Omega} \ll 1 \ , \nonumber \\
\dot{\psi}_x, \dot{\psi}_y &\sim& \sqrt{\xi\Omega} \ , \nonumber \\
p_x, p_y &\sim& \sqrt{\Omega/\xi} \ .
\ea
The first three relations reflect the kinematics of rotational
motion. The angles $\phi$ and $\theta$ range from zero to $2 \pi$ and
are, thus, of order unity. The right-hand side of Eq.~(\ref{moment})
shows that $r^2$ is the moment of inertia. By definition, the
rotational Hamiltonian is of order $\xi$. Thus, $r^2 \sim \xi^{-
  1}$. Writing the rotational Hamiltonian in terms of either
$\dot{\phi}$ and $\dot{\theta}$ or of $p_\phi$ and $p_\theta$ one
finds the second and the third of relations~(\ref{power}). The energy
of the vibrational modes is of order $\Omega$ and so are, therefore,
the ratios $\dot{\psi}_x / \psi_x$ and $\dot{\psi}_y / \psi_y$. The
occurrence of $\xi$ in the factors $\sqrt{\xi/\Omega}$ and
$\sqrt{\xi\Omega}$ is caused by the term that couples rotational and
vibrational motion. Relations three and six show that momenta scale
inversely to their corresponding coordinates. The inequality in
relation four expresses the assumption stated in Section~\ref{int}.
That assumption implies that the amplitudes of the vibrational modes
are small.

\subsubsection{Effective Hamiltonian to order $\xi$}

The leading terms describe vibrations and are of order $\Omega$. In
that order we have
\ba
H_\Omega &=& \int {\rm d}^3x\bigg\{ {p_x^2+p_y^2\over 2M} 
+ {M\over 2}\Big[ \omega_\parallel^2\left((\partial_3 \psi_x)^2 + (\partial_3 \psi_y)^2\right)\nonumber\\
&&+\omega_\perp^2
\left((\partial_1 \psi_x)^2 + (\partial_2 \psi_x)^2 + (\partial_1 \psi_y)^2 + (\partial_2 \psi_y)^2\right) 
\Big]\bigg\} \ .
\ea
The low-energy constants $M$, $\omega_\parallel$, and $\omega_\perp$
scale as $\Omega^{-1}$, $\Omega$, and $\Omega$, respectively. We note
that terms involving spatial derivatives of the NG fields play the
role of potential terms. Integration by parts yields
\ba
\label{HOmega}
H_\Omega &=& \int {\rm d}^3x\bigg\{ {p_x^2+p_y^2\over 2M} 
- {M\over 2}\Big[ \omega_\parallel^2\left(\psi_x\partial_3^2 \psi_x + \psi_y\partial_3^2 \psi_y \right)\nonumber\\
&&+\omega_\perp^2
\left(\psi_x(\partial_1^2 +\partial_2^2)\psi_x  + \psi_y(\partial_1^2 +\partial_2^2)\psi_y \right)
\Big]\bigg\} \ .
\ea
We decompose the NG fields
\ba
\label{psimodes}
\psi_x(\mathbf{x},t) = \sum_{\alpha} x_{\alpha}(t) \chi_\alpha(\mathbf{x})
\ , \
\psi_y(\mathbf{x},t) = \sum_{\alpha} y_{\alpha}(t) \chi_\alpha(\mathbf{x})
\ea 
into orthonormalized eigenmodes of the Helmholtz equation
\ba
\label{helmholtz}
-\left[\omega_\parallel^2\partial_3^2 +\omega_\perp^2 (\partial_1^2 +\partial_2^2)\right] 
\chi_{\alpha}(\mathbf{x}) 
= \omega_{\alpha}^2 \chi_{\alpha}(\mathbf{x}) \ .
\ea
Axial symmetry implies that many eigenfunctions are pairwise
degenerate in energy. We assume that the spectrum
$\omega_0<\omega_1\le\omega_2\le\ldots$ is ordered, and that the
lowest eigenvalue is not degenerate.  The momenta are decomposed
correspondingly,
\ba
\label{pmodes}
p_x(\mathbf{x},t) = \sum_{\alpha} p_{x;\alpha}(t) \chi_\alpha(\mathbf{x}) \ , \
p_y(\mathbf{x},t) = \sum_{\alpha} p_{y;\alpha}(t) \chi_\alpha(\mathbf{x}) \ .
\ea
The components $p_{x;\alpha}$ and $p_{y;\alpha}$ are canonical momenta
of the components $x_{\alpha}$ and $y_{\alpha}$, respectively. The
Hamiltonian~(\ref{HOmega}) becomes
\ba
H_\Omega &=& \sum_{\alpha} \left[ {p_{x;\alpha}^2+p_{y;\alpha}^2\over 2M} 
+ {M\over 2} \omega_{\alpha}^2\left(x_{\alpha}^2 +y_{\alpha}^2\right)\right] \ .
\ea
Calculation of the $\omega_\alpha$ would require a specific model for
the shape of the deformed nucleus and for the boundary conditions
imposed at the surface. That is of no interest here. In the EFT, only
the lowest-energy parameters $\omega_\alpha$ enter and need to be
adjusted to data.

The total Hamiltonian down to order $\xi$ is obtained by adding to
$H_\Omega$ the kinetic terms constructed in Section~\ref{kin}. That
gives
\ba
\label{Hxi}
H_\xi &=& H_\Omega + \frac{I^2 - K^2}{2 r^2} \ .
\ea
Vibrational and rotational motion are coupled via the gauge term in
Eq.~(\ref{moment}).

For the construction of invariants at higher order, we note that it
might be useful to consider the vectors
\ba
\vec{r}_\alpha &\equiv& x_{\alpha}\vec{e}_\theta + y_{\alpha}\vec{e}_\phi \nonumber\\ 
\vec{p}_\alpha &\equiv& p_{x;\alpha}\vec{e}_\theta + p_{y;\alpha}\vec{e}_\phi 
\ea
that are in the tangent plane. As an example, we note that the
intrinsic angular momentum~(\ref{K}) can be written as
\be
K\vec{e}_r = \sum_\alpha \vec{r}_\alpha\times\vec{p}_\alpha \ .
\ee

\subsection{Quantization}

For the vibrational momenta, the quantization rules are standard,
\ba
p_{x;\alpha} &=& -i\partial_{x_{\alpha}} \ , \nonumber \\
p_{y;\alpha} &=& -i\partial_{y_{\alpha}} \ ,
\label{quantization}
\ea
and the resulting spectrum is that of infinitely many uncoupled
two-dimensional harmonic oscillators with an $SO(2)$ symmetry and
frequencies $\omega_{\alpha}$. The spectrum of each $SO(2)$ symmetric
oscillator is 
\be
E_\alpha(n_\alpha,k_\alpha) = \omega_\alpha(2n_\alpha+|k_\alpha|) \ .
\ee
Here $n_a$ with $n_\alpha=0,1,2,\ldots$ is the principal quantum
number and $k_a$ with $k_\alpha=0,\pm1,\pm2,\ldots$ is the projection
of the angular momentum of the vibrational modes. We neglect
zero-point energies. We have
\ba
K &=& \sum_{\alpha} \left( x_{\alpha} p_{y; \alpha} - y_{\alpha} p_{x; \alpha}
\right) \ .
\ea
The eigenvalue $k$ of the operator $K$ depends upon the eigenstate of the
Hamiltonian that $K$ is acting on and is given by
\be
k = \sum_\alpha k_\alpha \ .
\ee
The quantization of angular momentum is standard. In
Eq.~(\ref{moment}) we replace the operator $K$ by its eigenvalue
$k$. The eigenfunctions of the rotational Hamiltonian obtained by
that replacement are Wigner $D$ functions~\cite{varshalovich1988},
\ba
&& \frac{\vec{I}^2 - k^2}{2 r^2} D_{mk}^I(\phi, \theta, 0)
 = {I(I+1) - k^2 \over 2 r^2} D_{mk}^I(\phi, \theta, 0) \ ,
\ea
with $I = |k|, |k|+1, |k|+2, \ldots$. The moment of inertia is
$r^2$. Due to ${\cal R}$ parity, only linear combinations of wave
functions $D_{mk}^I+(-1)^kD_{m-k}^I$ are admissible. The spectrum
consists of rotational bands on top of vibrational band heads with
integer spin $|k|$.

\subsection{Higher-order terms}

The amplitudes of the surface vibrations are small and of order $\ve
\sim \Omega / \Lambda$. Terms of next order contain higher powers of
these amplitudes and are, thus, of order $\Omega \ve$. Relevant
kinetic invariants are then
\ba
\label{kininv}
&& \int {\rm d}^3x  \ (\psi_x p_y - \psi_y p_x)^2 \ , \nonumber \\
&& \int {\rm d}^3x  \ (\psi^2_x + \psi^2_y) (p^2_x + p^2_y) \ , 
\nonumber \\
&& \frac{\cos^2 \theta}{\sin^2 \theta} \int {\rm d}^3x \ (\psi_x p_y-
\psi_y p_x)^2 \ .
\ea
The expansion coefficients of $p_x, p_y$ are quantized as in
Eqs.~(\ref{quantization}). The first two terms are constructed from
invariants in a very obvious way.  The last term results from the
coupling of rotational motion and vibrations. As shown by the last
term in the covariant derivative in Eq.~(\ref{Dt}), that coupling is
proportional to $\cos \theta$ and yields the factor $\cos^2
\theta$. The expression~(\ref{vecv}) for the velocity shows that
replacing $\dot{\phi}$ by the conjugate momentum $p_\phi$ produces a
factor $(\sin\theta)^{- 1}$, hence the factor $(\sin\theta)^{- 2}$.
Explicitly we may use Eq.~(\ref{Ialt}) to write $\int {\rm d}^3 x
\vec{I}^2$ as
\ba
\int {\rm d}^3 x \left[\vec{r} \times \left(\vec{p}-{(\psi_xp_y -\psi_y p_x)\cot\theta\over r} \vec{e}_\phi\right) + (\psi_x p_y -\psi_y p_x)\vec{e}_r \right]^2 \ .
\ea
The integration extends only over the arguments of $\psi_x, \psi_y,
p_x, p_y$. The square of the term proportional to $\vec{e}_\phi$
yields the third expression~(\ref{kininv}). An alternative view on the
occurrence of the term proportional to $\cot^2\theta$ is given in the
\ref{app}.

Expanding the first expression~(\ref{kininv}) into eigenmodes yields
\be
\sum_{\alpha\beta\gamma\delta}\chi_{\alpha\beta\gamma\delta} (x_\alpha p_{y;\beta}-y_\alpha p_{x;\beta}) (x_\gamma p_{y;\delta}-y_\gamma p_{x;\delta})
\ee
with
\be
\label{chi4} 
\chi_{\alpha\beta\gamma\delta}\equiv \int {\rm d}^3 x
\chi_\alpha(\mathbf{x})\chi_\beta(\mathbf{x})\chi_\gamma(\mathbf{x})\chi_\delta(\mathbf{x}) \ .  
\ee 
Because of the axial symmetry of the eigenfunctions
$\chi_\mu(\mathbf{x})$, the expression
$\chi_{\alpha\beta\gamma\delta}$ vanishes unless the azimuthal quantum
numbers contained in the labels ($\alpha,\beta,\gamma,\delta$) sum up
to zero. Thus, we can rewrite the first two invariants in
Eq.~(\ref{kininv}) as
\ba
\sum_{\alpha\beta\gamma\delta}\chi_{\alpha\beta\gamma\delta} 
\left(\vec{r}_\alpha\times\vec{p}_\beta\right)\cdot
\left(\vec{r}_\gamma\times\vec{p}_\delta\right) \ , \nonumber\\
\sum_{\alpha\beta\gamma\delta}\chi_{\alpha\beta\gamma\delta} 
\left(\vec{r}_\alpha\cdot\vec{r}_\beta\right)
\left(\vec{p}_\gamma\cdot\vec{p}_\delta\right) \ .
\ea

Potential terms in the Hamiltonian involve invariants constructed from
the spatial derivatives of the fields $\psi_x$ and $\psi_y$. Examples are
\ba
\int {\rm d}^3 x \ (\psi_x \mathbf{\nabla} \psi_y - \psi_y \mathbf{\nabla} \psi_x)^2
\ , \nonumber \\
\int {\rm d}^3 x \ (x^2 + y^2) ( (\mathbf{\nabla}x)^2 + (\mathbf{\nabla} y)^2 ) \ .
\ea
Here $\mathbf{\nabla}$ stands for the gradient with respect to
$\mathbf{x}$ and acts onto the modes~(\ref{psimodes}). Formally, such
terms can be written as
\ba
\sum_{\alpha\beta\gamma\delta}\tilde{\chi}_{\alpha\beta\gamma\delta} 
\left(\vec{r}_\alpha\times\vec{r}_\beta\right)\cdot
\left(\vec{r}_\gamma\times\vec{r}_\delta\right) \ , \nonumber\\
\sum_{\alpha\beta\gamma\delta}\tilde{\chi}_{\alpha\beta\gamma\delta} 
\left(\vec{r}_\alpha\cdot\vec{r}_\beta\right)
\left(\vec{r}_\gamma\cdot\vec{r}_\delta\right) \ .
\ea
Here, $\tilde{\chi}_{\alpha\beta\gamma\delta}$ results from an
integral similar to Eq.~(\ref{chi4}) but with derivatives in the
integrand.

\section{Results}
\label{results}

In this Section we review some of the results obtained with the EFT
approach to deformed nuclei. First, we discuss the spectrum of the
Hamiltonian~(\ref{Hxi}). Second, we briefly discuss the coupling of
the EFT degrees of freedom to electromagnetic fields and some
electromagnetic transitions. Third, we consider how the EFT
constructed in this paper reduces to an effective theory when the
number of NG modes is truncated to a (small) set. We end with a brief
discussion of common features of and differences between the EFT and
collective models.

\subsection{Spectra}
\label{sub:spectra}

Our EFT applies equally to low-lying states in axially symmetric
nuclei and in axially symmetric molecules. In both cases, the ground
state has spin $k = 0$, and the ground-state rotational band has spins $I =
0, 2, 4, \ldots$. The results differ for the next states in the
spectrum, however, because nuclei possess positive ${\cal R}$
parity. That excludes a positive-parity state with $k = 1$. Such a
state requires the breaking of a Cooper pair~\cite{heyde2010} since
two identical fermions in a single $j$ shell can only couple to even
angular momenta. Although our EFT approach does not include fermionic
degrees of freedom, the effects of pairing are seen to be indirectly
taken into account. In axially symmetric molecules, the lowest
vibrational state has a single quantum in the $\alpha = 0$ mode,
quantum numbers $k = k_0 = 1$~\cite{herzberg1945}, excitation energy
$\omega_0$, and negative ${\cal R}$ parity. In nuclei, that state is
excluded. The lowest-lying vibrational states have quantum numbers
$(n_0 = 1,k_0 = 0)$ and $(n_0 = 0,|k_0| = 2)$. For the leading-order
Hamiltonian, these are degenerate at energy $2 \omega_0$. The first of
these ($k = 0$) is commonly referred to as the ``$\beta$'' vibration.
The second ($k = 2$) is the ``$\gamma$'' vibration. The degeneracy is
lifted by terms of higher order in the Hamiltonian. In most deformed
nuclei there is indeed a quasi-degenerate doublet of excited
vibrational bands for which the difference in band-head energy is much
smaller than the excitation energy $2 \omega_0$. Beyond the
ground-state band and the nearly degenerate $k = 0$ and $k = 2$
excited bands, the vibrational spectra of nuclei become
non-universal. Details depend on the precise values of the energies
$\omega_0 < \omega_1 \le \omega_2 \le \ldots$, and different nuclei
are expected to exhibit different vibrational spectra. Within the
leading-order Hamiltonian, all rotational bands have identical
rotational constants (or moments of inertia). Differences in
rotational constants are caused by terms of higher order in the power
counting~\cite{zhang2013}.

It is of interest to compare these theoretical results with the
extensive data for $^{168}$Er~\cite{davidson1981} and
$^{162}$Dy~\cite{aprahamian2006}. In $^{168}$Er, the head of the $k =
2$ band is at 821~keV, that of the excited $k = 0$ band is at
1217~keV. Thus, $2 \omega_0 \approx 1$~MeV. The splitting between the
two states is about 40\% of $2 \omega_0$. The rotational excitation
energies are less than $\xi \approx 80$~keV so that $\xi / \Omega
\approx 1 / 10$. The moments of inertia for the ground-state band and
for the excited $k = 2$ and $k = 0$ bands are $r^{-2} \approx 27$~keV,
$r^{-2} \approx 25$~keV, and $r^{-2} \approx 20$~keV, respectively.
The lowest negative-parity state has $k^\pi = 0^-$ and is at about
1100~keV, while the lowest $k^\pi = 1^-$ state is at about 1360~keV.

For $^{162}$Dy, the excited $k = 2$ vibrational state is at about
888~keV, the lowest $k = 0$ excited state is at about 1400~keV. Thus
$2\omega_0\approx 1.1$~MeV, and the splitting of the $k = 0$ and $k =
2$ states amounts to almost 50\% of $2\omega_0$. The moments of
inertia of the ground-state band, the $k = 2$, and the $k = 0$ bands
are $r^{-2}\approx 27$~keV, $r^{-2}\approx 25$~keV, and $r^{-2}\approx
18$~keV, respectively.

\subsection{Coupling to electromagnetic fields}

In this Subsection we fill a gap. In the main part of the paper we
have not addressed the coupling of the EFT degrees of freedom to
electromagnetic fields. We do so now and work in the Coulomb
gauge. Then we only need to consider the vector potential
$\vec{A}(\vec{r})$. The gauging of the rotational degrees of freedom
$(\theta, \phi)$ is straightforward and has been discussed in detail
in Ref.~\cite{coello2015}. One finds that expression~(\ref{pvec}) is
changed into
\be
\vec{p} \to \vec{p}-q\vec{A} \ .
\ee
Here, the charge $q$ is a LEC and can be adjusted to data using a
single transition within the ground-state band.

We turn to the electromagnetic coupling of the NG fields. We recall
that for every point $\mathbf{x}$, the fields $\psi_x(\mathbf{x})$ and
$\psi_y(\mathbf{x})$ ``live'' in the tangential plane of the
two-sphere at $\vec{r}$. We also recall the
expansions~(\ref{psimodes}) and (\ref{pmodes}) for the fields and the
associated momenta. The latter, given by
\be
\vec{p}_\alpha \equiv p_{x;\alpha}\vec{e}_\theta + p_{y;\alpha}\vec{e}_\phi 
\ee
are vectors in the tangential plane. Thus we can gauge them as 
\be
\vec{p}_\alpha \to \vec{p}_\alpha -q_\alpha\vec{A} \ .
\ee
Here, $q_\alpha$ is a LEC that can be adjusted to data by means of a
single inter-band transition from the rotational band with the
vibrational band head at an excitation energy $2\omega_\alpha$ to the
ground-state band. 

Why do the charges $q_\alpha$ depend on the mode $\chi_\alpha$? The
eigenmodes $\chi_\alpha$ solve the Helmholtz
equation~(\ref{helmholtz}). They differ from each other. The effective
charge each mode carries is specific to that mode. As is the case for
the energies $\omega_\alpha$, a microscopic calculation of that charge
would require a model for the shape of the deformed nucleus and for
the boundary conditions imposed at the surface.

\subsection{From the EFT to effective theories}

The dynamical variables in our EFT for deformed nuclei are the
rotational degrees of freedom $(\theta, \phi)$ and the amplitudes
$(x_\alpha, y_\alpha)$ of the NG modes. Both the excitation energies
$\omega_\alpha$ and the charges $q_\alpha$ are mode-specific
parameters. The number of parameters increases with the number of
modes considered. It is, therefore, tempting to confine attention to
the (few) modes that are below the breakdown energy $\Lambda$ of the
EFT. That truncation reduces the EFT to an effective theory. In
references~\cite{papenbrock2011,zhang2013,coello2015}, aspects of such
an effective theory were investigated and used. We discuss the
relation between that effective theory and the present EFT.

For simplicity we restrict the EFT developed in
Sect.~\ref{construction} to the lowest excitations with energy $2
\omega_0$ as discussed in Subsection~\ref{sub:spectra}. At order
${\cal O}(\Omega)$ we deal with two two-dimensional isotropic harmonic
oscillators. At order ${\cal O}(\xi)$ the oscillators are coupled to a
rigid rotor. That yields two rotational bands with identical moments
of inertia on top of both, the $k = 0$ band head and the degenerate
$|k| = 2$ band head.

In contradistinction, the effective
theory~\cite{papenbrock2011,zhang2013,coello2015} starts from the
emergent symmetry breaking of five quadrupole degrees of freedom
$d_\mu$. The components $d_{\pm 1}$ are replaced by two rotation
angles $(\theta, \phi)$ and become the modes with excitation energy
$\xi$. The component $d_0$ and the components $d_{\pm 2}$ have the
energy scale $\Omega$. At order ${\cal O}(\Omega)$ the Hamiltonian is
a three-dimensional axially symmetric harmonic oscillator with
frequency $\tilde{\omega}_0$ for the $d_0$ degree of freedom and
frequency $\tilde{\omega}_2$ for the $d_{\pm 2}$ degrees of freedom.
The frequencies $\tilde{\omega}_0$ and $\tilde{\omega}_2$ have
magnitude $\Omega$ but are not necessarily equal to each other. The
lowest vibrational excitations have energy $\tilde{\omega}_0$ (one
quantum in the $d_0$ mode) or energy $\tilde{\omega}_2$ (one angular
excitation of the two-dimensional isotropic oscillator in the $d_{\pm
  2}$ modes). At order ${\cal \xi}$ these oscillators are coupled to
the rigid rotor with the degrees of freedom $(\theta, \phi)$. That
yields a rotational band with $k = 0$ on top of the $d_0$ excitation
and another one with $k = 2$ on top of the $d_{\pm 2}$ excitation.
Thus, at order $\xi$ the spectra of the effective theory and of the EFT
differ from each other because the former has nondegenerate
frequencies $\tilde{\omega}_0 \ne \tilde{\omega}_2$ while the latter
has equal frequencies $2 \omega_0$ for the $k = 0$ and $|k| = 2$
bands. In the effective theory, the difference in frequencies
$\tilde{\omega}_0 - \tilde{\omega}_2$ is expected to be small compared
to $\Omega$ and is viewed as a higher-order correction. In that sense
the effective theory and the EFT are equivalent low-energy theories.

The EFT for deformed nuclei yields a systematic approach that is based
on symmetry principles alone. We mention two examples that show how
within that approach, small deviations between data and the
traditional collective models can be understood and addressed. These
are the variation of the moment of inertia with the vibrational band
head~\cite{zhang2013}, and the weak inter-band $E2$
transitions~\cite{coello2015}. We do so in the framework of the
effective theory which, as just pointed out, is a simplified version
of the EFT.

In nuclei, the moments of inertia (or rotational constants) of
rotational bands on top of different vibrational excitations differ by
relatively small amounts. Examples were given in
Subsection~\ref{sub:spectra}. In the effective theory, rotational
constants at order $\xi$ have magnitude $\xi^{-1}$ and for different
rotational bands are equal. Taking account of higher-order corrections
of relative order $\xi/\Omega$, consistency with the data in
$^{166,168}$Er and $^{232}$Th is attained~\cite{zhang2013}.

For $E2$ transitions the effective theory predicts that intra-band
transitions are strong and that inter-band transitions are suppressed
by a factor of order $\xi/\Omega$. Gauging of the effective theory
shows that the inter-band transitions are governed by two additional
LECs. These parameters do not appear in the traditional collective
models, causing the latter to overpredict the faint inter-band
transition strengths by factors 2--10~\cite{rowe2009}. The effective
theory remedies this problem~\cite{coello2015} and thereby offers a
solution to a long-standing discrepancy.  Nuclei are non-rigid
rotors. Within the EFT it becomes clear, for instance, that deviations
from rigid-rotor expectations for the spectrum and quadrupole
transitions are similar in relative size. 

These examples show that a systematic and controlled approach to
nuclear deformation is possible, and that small but significant
deviations between data and collective models can be understood and
addressed. Within the EFT such problems are not treated by simply
adding terms with additional fit parameters. Instead, arguments of
symmetry alone are used to determine which corrections arise at each
order of power counting. The procedure is unambiguous. Consistency of
the EFT approach requires that LECs are of natural size, i.e., have a
magnitude that is in agreement with expectations from the power
counting.  The procedure also shows which additional assumptions are
made when correction terms are used in the traditional collective
models.

The expectation that LECs are of natural size leads to simple
estimates of the theoretical uncertainties at any order of the power
counting.  For deformed nuclei, such uncertainty estimates for $B(E2)$
transitions suggest, for instance, that it might be profitable to
remeasure or re-evaluate data for certain intra-band
transitions~\cite{coello2015}. Most interestingly, the EFT approach
can also be used to truly quantify uncertainties. Assumptions about
the natural size and distribution of LECs can be quantified as priors,
and Bayesian statistics can be used to quantify theoretical
uncertainties as degree-of-belief intervals that have a statistical
meaning~\cite{schindler2009,cacciari2011,bagnaschi2015,furnstahl2014c,furnstahl2015,coello2015b}.

\section{Summary}
\label{summary}

We have reviewed the EFT approach to nuclei with intrinsically
deformed but axially symmetric ground states. The deformation is
viewed as a case of emergent symmetry breaking, the analogue of
spontaneous symmetry breaking in infinite systems. Accordingly, our
EFT extends well-known approaches to spontaneous symmetry breaking in
infinite systems to emergent symmetry breaking characteristic of
finite systems. That is done using, in addition to the familiar
Nambu-Goldstone modes, additional modes that account for nuclear
rotation. The Hamiltonian consists of invariants that are constructed
from the said modes using symmetry arguments alone. Each invariant is
multiplied by a constant that has to be fitted to data. The invariants
are ordered by power-counting arguments. In leading order, the
Hamiltonian describes rotations and vibrations, each vibrational state
serving as band head of a rotational band. Terms of higher order allow
for a systematic improvement and lift degeneracies. That Hamiltonian
governs the spectra of deformed nuclei at low excitation energies. The
construction leaves no room for guess work: Each invariant and its
order are well defined. These facts make it possible to address small
but significant differences between data and traditional collective
models. Examples are changes of the moments of inertia as the band
head changes, or the magnitude of the faint inter-band $E2$
transitions.

The Nambu-Goldstone modes are defined in the coset space $SO(3) /
SO(2)$. The resulting non-linear realization of symmetry breaking is
at the heart of the EFT. Upon quantization the Nambu-Goldstone modes
give rise to the nuclear vibrational modes. Very recently, low-energy
vibrations in spherical nuclei have also been approached in an
EFT~\cite{coello2015b}. Unlike the present approach, the EFT for such
nuclear vibrations is based on the usual linear realization of
rotational symmetry. That approach suggests that certain isotopes of
Ni, Ru, Pd, Cd, and Te can be viewed as anharmonic quadrupole
oscillators. The approach describes low-lying spectra and
electromagnetic properties consistently within quantified theoretical
uncertainties.

The EFT approach to heavy nuclei can be extended in several
directions. Most interesting is probably the coupling of fermionic
degrees of freedom to the bosonic fields discussed in this paper. That
approach -- similar in spirit to halo EFT -- could open the way
towards a model-independent theory of odd-mass and odd-odd heavy
nuclei. Another possibility is an extension of our parity-conserving
EFT to emergent parity breaking~\cite{butler1996}.  Octupole
excitations, i.e., states with spin/parity $I^\pi=1^-,2^-,3^-,\ldots$
are low-lying vibrations in rare-earth nuclei and the lowest-lying
vibrations in the actinides.

Ultimately, we wish to better understand how collective modes arise in
complex nuclei, and to reliably quantify theoretical
uncertainties. EFT approaches to heavy nuclei have the potential to
deliver both.

\ack One of us (HAW) acknowledges support by the Simons Center for
Geometry and Physics at Stony Brook University where part of this
paper was written.  TP's work is supported in part by the U.S.
Department of Energy, Office of Science, Office of Nuclear Physics,
under award No. DE-FG02-96ER40963 (University of Tennessee), and under
contract No. DEAC05-00OR22725 (Oak Ridge National Laboratory).

\appendix
\section{Transformation properties under rotations}
\setcounter{section}{1}
\label{app}
In this Appendix, we derive the transformation properties of the
degrees of freedom employed in the EFT. We also use Noether's theorem
for the derivation of angular momentum as the conserved quantity, and
show that the monopole gauge fields change under rotations by a total
gradient.

A rotation $\delta \vec{\alpha} = (\delta \alpha_x, \delta \alpha_y,
\delta \alpha_z)$ about infinitesimal angles $\delta \alpha_k$ around
the laboratory $k = x, y, z$ axes changes the radial unit vector
$\vec{e}_r$ and the unit vectors $\vec{e}_\theta$ and $\vec{e}_\phi$
in the tangential plane at the point $(\theta, \phi)$ as
\ba
\label{app1}
\vec{e}_j (\theta, \phi) &\to& \vec{e}_j (\theta, \phi) + \delta
\vec{\alpha} \times \vec{e}_j (\theta, \phi) \ .
\ea
Here $j = r, \theta, \phi$. The rotation $\delta \vec{\alpha}$ also
changes the point $(\theta, \phi)$ on the sphere to $(\theta + \delta
\theta, \phi + \delta \phi)$, and the three unit vectors at this
rotated point are
\ba
\label{app2}
\vec{e}_r (\theta + \delta \theta, \phi + \delta\phi) &=& \vec{e}_r
(\theta, \phi) + \delta \theta \vec{e}_\theta (\theta, \phi) +
\delta \phi \sin \theta \vec{e}_\phi (\theta, \phi) \ , \nonumber \\
\vec{e}_\theta (\theta + \delta \theta, \phi + \delta \phi) &=&
\vec{e}_\theta (\theta, \phi) - \delta \theta \vec{e}_r (\theta, \phi)
+ \delta \phi \cos \theta \vec{e}_\phi (\theta, \phi) \ , \nonumber \\
\vec{e}_\phi (\theta + \delta \theta, \phi + \delta \phi) &=&
\vec{e}_\phi (\theta, \phi) - \delta \phi \left[ \sin \theta
\vec{e}_r (\theta, \phi) + \cos \theta \vec{e}_\theta (\theta,
\phi) \right] \ .
\ea
Equating the expressions on the right-hand sides of Eq.~(\ref{app1})
for $j = r$ and of the first of Eqs.~(\ref{app2}) yields
\ba
\label{trafo_theta_phi}
\left(
\begin{array}{c}
\delta\theta\\
\delta\phi
\end{array}\right) 
=\left[
\begin{array}{ccc}
-\sin \phi & \cos \phi & 0\\
-\cos \phi \cot \theta & - \sin \phi \cot \theta & 1
\end{array}\right] 
\left(
\begin{array}{c}
\delta \alpha_x\\
\delta \alpha_y\\
\delta \alpha_z
\end{array}\right) . 
\ea
That expression shows that the rotation $\delta \vec{\alpha}$ induces
a non-linear transformation of the angles $(\theta, \phi)$. 
It follows
that a scalar function $f(\theta, \phi)$ changes under the rotation
$\delta \vec{\alpha}$ as
\be
f(\theta - \delta \theta, \phi - \delta \phi)
= f(\theta, \phi) - \delta \theta \partial_\theta f(\theta, \phi)
- \delta \phi \partial_\phi f(\theta, \phi) \to f(\theta, \phi) \ ,
\ee
with $\delta \theta$ and $\delta \phi$ given by
Eq.~(\ref{trafo_theta_phi}).

We consider a vector $\vec{a} = a_\theta \vec{e}_\theta(\theta, \phi)
+ a_\phi \vec{e}_\phi(\theta, \phi)$ in the tangential plane at the
point $(\theta, \phi)$. Under the rotation $\delta \vec{\alpha}$, a
radial vector is transformed into a radial vector and a tangential
vector into a tangential vector. However, the rotated pair
$\vec{e}_\theta$, $\vec{e}_\phi$ of tangential vectors does not
coincide with the corresponding pair of tangential basis vectors at
the point reached by the rotation. In general, the two pairs differ by
a rotation in the tangential plane. We now determine the infinitesimal
value of angle of that rotation.

Under a rotation $\delta \vec{\alpha}$, the basis vectors in the
tangential plane transform as in Eq.~(\ref{app1}) for $j = \theta,
\phi$. It is straightforward to take scalar products of these vectors
with the basis vectors $\vec{e}_\theta(\theta + \delta \theta, \phi +
\delta \phi)$ and $\vec{e}_\phi(\theta + \delta \theta, \phi + \delta
\phi)$ in the tangential plane at $(\theta + \delta \theta, \phi +
\delta \phi)$ as defined in Eqs.~(\ref{app2}). Alternatively we might
use all three Eqs.~(\ref{app2}), solve for the basis vectors at
$(\theta + \delta \theta, \phi + \delta \phi)$, re-express the rotated
tangent vectors~(\ref{app1}) in terms of the latter, and compute the
scalar products. In either case we obtain the transformation law for
the basis vectors in the tangential plane
\ba
\left(\begin{array}{c}
\vec{e}_\theta\\
\vec{e}_\phi
\end{array}\right)
\to\left[\begin{array}{cc}
1 & \delta\gamma \\
-\delta\gamma & 1\end{array}\right]
\left(\begin{array}{c}
\vec{e}_\theta\\
\vec{e}_\phi
\end{array}\right) \ .
\ea
Here
\be
\label{gamma}
\delta\gamma = {\cos \phi \over \sin \theta} \delta \alpha_x + {\sin
\phi \over \sin \theta} \delta \alpha_y \ .
\ee
For the components of the vector $\vec{a} = a_\theta \vec{e}_\theta(\theta,
\phi) + a_\phi \vec{e}_\phi(\theta, \phi)$ we obtain correspondingly
\ba
\label{trafo_tangent}
\left(\begin{array}{c}
a_\theta\\
a_\phi
\end{array}\right)
\to\left[\begin{array}{cc}
1 & -\delta\gamma \\
\delta\gamma & 1\end{array}\right]
\left(\begin{array}{c}
a_\theta\\
a_\phi
\end{array}\right) \ .
\ea
A rotation $\delta \vec{\alpha}$ induces a rotation of vectors in the
tangential plane by the angle $\delta \gamma$. That is an alternative
explanation of why the NG fields (which ``live'' in the tangential
plane) transform linearly under rotations albeit with a complex angle.
Using Eqs.~(\ref{trafo_tangent}) and (\ref{gamma}) yields the
transformation properties of the NG fields
\ba
\label{trafo_psi}
\left(
\begin{array}{c}
\delta\psi_x\\
\delta\psi_y
\end{array}\right) 
=\left[
\begin{array}{ccc}
-\psi_y{\cos \phi\over\sin\theta} & -\psi_y {\sin\phi \over \sin\theta}& 0\\
\psi_x{\cos \phi\over\sin\theta} & \psi_x {\sin\phi \over \sin\theta}& 0
\end{array}\right] 
\left(
\begin{array}{c}
\delta \alpha_x\\
\delta \alpha_y\\
\delta \alpha_z
\end{array}\right) . 
\ea
Employing the matrix elements in Eqs.~(\ref{trafo_theta_phi}) and (\ref{trafo_psi}) 
together with Noether's theorem yields the angular-momentum components
\ba
I_x &=& -p_\theta\sin\phi -p_\phi \cos\phi\cot\theta + {\cos\phi\over\sin\theta}\int{\rm d}^3 x\left(\psi_x p_y - \psi_y p_x\right) \ ,\nonumber\\
I_y &=& p_\theta \cos\phi -p_\phi \sin\phi\cot\theta + {\sin\phi\over\sin\theta}\int{\rm d}^3 x\left(\psi_x p_y - \psi_y p_x\right) \ ,\\
I_z&=&p_\phi\nonumber
\ea
as the conserved quantities. This is Eq.~(\ref{I}). 

We apply these results to the vector potential~(\ref{Amono}). Under the
rotation $\delta \vec{\alpha}$ we have
\ba
{\cot\theta\over r}\vec{e}_\phi &\to& {\cot\theta\over r}\vec{e}_\phi -\delta\gamma{\cot\theta\over r}\vec{e}_\theta +{\delta\theta\over r\sin^2\theta}\vec{e}_\phi\nonumber\\
&=& {\cot\theta\over r}\vec{e}_\phi +\vec{\nabla}\left(\delta\gamma\right) .
\ea
The rotation changes the monopole vector potential by the gradient of
a scalar function, i.e., by an unobservable gauge
transformation. Hence, any combination
\be
\left(\vec{p} - C\cot\theta\vec{e}_\phi\right)^2
\ee
with a rotational scalar $C$ is invariant under rotations. That explains
the occurrence of terms involving $\cot^2 \theta$ in the higher-order
corrections of the EFT.

\section*{References}


\providecommand{\newblock}{}

\end{document}